\documentclass[12pt]{iopart} 

\usepackage{graphicx}
\usepackage{comment}

\begin{document}

\title{System size dependence of hadron $p_T$ spectra in $p+p$ 
and Au+Au collisions at $\sqrt{s_{NN}}$ = 200 GeV}

\author{P. K. Khandai$^1$, P. Sett$^{2}$, P. Shukla$^{2}$, V. Singh$^1$}

\address{$^1$Department of Physics, Banaras Hindu University, Varanasi 221005, India} 
\address{$^2$Nuclear Physics Division, Bhabha Atomic Research Center, Mumbai 400085, India}
\ead{pshukla@barc.gov.in}

\begin{abstract}
  We make a systematic study of transverse momentum ($p_T$) spectra of hadrons produced 
in $p+p$ and in different centralities of Au+Au collisions at $\sqrt{s}=200$\,GeV using 
phenomenological fit functions. The Tsallis distribution gives a very good description of 
hadron spectra in $p+p$ collisions with just two parameters but does not produce 
the same in Au+Au collisions at intermediate $p_{T}$.   
  To explain the hadron spectra in heavy ion collisions in a wider $p_{T}$ range we propose a 
modified Tsallis function by introducing an additional parameter which accounts for 
collective flow. 
The new analytical function gives a very good description of both mesons and baryons 
spectra at all centralities of Au+Au collisions in terms of parameters having a  
potential physics interpretation. 
  With this modified Tsallis function we study the spectra of pions, kaons, protons, $\Lambda$
and $\Xi^{-}$ as a function of system size at $\sqrt{s_{NN}}$ = 200 GeV. 
  The parameter representing transverse flow increases with system size and is found to be 
more far baryons than that for mesons in central Au+Au collisions.
  The freeze-out temperature for baryons increases with centrality in Au+Au collisions more 
rapidly than for mesons. 
\vspace{0.6in}  
\end{abstract}


\noindent{\it QGP, Tsallis distribution, hadron spectra \/}

\submitto{\JPG}


\maketitle

\section{Introduction}
  The heavy ion collisions at relativistic energies are carried out to study the properties of 
strongly interacting matter at high temperature, where a phase transition to a Quark Gluon Plasma (QGP) 
is expected.
  Measurements in Au+Au collisions, performed  at the Relativistic Heavy Ion Collider (RHIC)
already point to the formation of Quark Gluon Plasma (QGP) \cite{WP1,WP2,HADPHEN}. 
Experiments at RHIC continue to study the detailed properties of the strongly interacting matter using $p+p$ and Au+Au 
systems at colliding energies with $\sqrt{s_{NN}}$ ranging from 7.7 GeV to 200 GeV.
  The $p+p$ collisions are used as baseline and are important to understand the particle production 
mechanism \cite{PPPROD}.
 Transverse momentum ($p_T$) distributions of identified hadrons are the most common tools used 
to study the dynamics of high energy collisions. 
  The high $p_T$ hadrons are important for QGP studies as they measure the jet quenching \cite{jet_quenching} 
effect in QGP. The low $p_T$ hadrons arise from multiple 
scatterings and follow an exponential distribution suggesting particle production in a 
thermal system \cite{Gatoff_Wong_PRD46_1992}. In addition, the hadron spectra at 
intermediate $p_T$ are sensitive to effects arising from quark recombination \cite{recommodel} 
in heavy-ion collisions.

  The Tsallis distribution  \cite{Tsallis, scale_ref} gives an excellent description of $p_{T}$ spectra 
 of all identified mesons measured in $p+p$ collisions at $\sqrt{s}=200$\,GeV \cite{PPG099}.
 It interprets the system in terms of temperature and a parameter which 
measures temperature fluctuation. 
  In a recent work \cite{IJMPA}, we used the Tsallis distribution to describe the $p_T$ spectra 
 of identified charged hadrons measured in $p+p$ collisions at RHIC ($\sqrt{s}$ = 62.4, 200 GeV) and 
 at LHC ($\sqrt{s}$ = 0.9, 2.76 and 7.0 TeV) energies. It has been shown \cite{scale_ref,IJMPA} that the 
functional form of the Tsallis distribution with thermodynamic origin is of the same form as the QCD-inspired 
Hagedorn distribution \cite{HAG1, HAGEFACT}.
 The non-extensive parameter $q$ of Tsallis is thus related to the power $n$ of the Hagedorn function.  
 
 There have been many attempts to use the Tsallis or Hagedorn distributions to fit the $p_{T}$ spectra 
of hadrons produced in heavy ion collisions. To fit the $\pi^{0}$ spectrum measured in Au+Au collisions,  
the PHENIX collaboration used a function referred as the modified Hagedorn formula \cite{PPG088, PPG077}. 
 The spectra of other mesons are then obtained by $m_{T}$ scaling which were used to get the hadronic 
 decay cocktails for single or dielectron spectra.
  In one of our works \cite{mesonscaling}, we used this modified Hagedorn formula to test the  
$m_{T}$ scaling for mesons extensively at $\sqrt{s}$=200\,GeV and for baryons in the 
work  \cite{baryonscaling}. While the modified Hagedorn formula gives a very good description of
hadron $p_T$ distributions, its parameters lack a physics interpretation. 
 There have been few other attempts to find a good fit function for hadron spectra measured in
various colliding systems e.g. work in Ref. \cite{Bylinkin_arxiv:1008.0332(2010)} 
adds another exponential term with Tsallis function to fit the data.

  The shape of the $p_T$ distributions of mesons and baryons resemble the $p_T$ distribution of 
quarks as shown by recombination models \cite{Greco}. Thus, the change in the $p_T$ spectra of 
quarks due to collective flow etc. will be reflected in the measured $p_T$ distribution of 
hadrons in heavy ion collisions which could be included in phenomenological fit functions 
such as the Hagedorn or Tsallis formula. 
 To explain the hadron spectra in heavy ion collisions in the large $p_{T}$ range, we propose a 
modified Tsallis function by introducing an additional parameter which accounts for transverse flow. 
The new analytical function gives very good description of both mesons and baryons 
spectra at all centralities of Au+Au collisions in terms of parameters having
potential physics interpretation. 
 We make a systematic study of the parameters of the modified Tsallis function 
for pions, kaons, protons, $\Lambda(1115)$, and $\Xi^{-}$ as a function of system size 
at $\sqrt{s_{NN}}$ = 200 GeV using measured hadron spectra from PHENIX and STAR experiments.

\section{The Tsallis distribution and collective transverse flow}
  The Tsallis distribution \cite{Tsallis,scale_ref}, describes a thermal system 
in terms of two parameters $T$ and $q$, is given by
\begin{equation}
 E\frac{d^3N}{dp^3}  =  C_q \left(1+(q-1)\frac{E}{T}\right)^{-1/(q-1)}.
\label{Tsallis_eqn}
\end{equation}
  Here $C_{q}$ is the normalization constant, $E$ is the particle energy, $T$ is the temperature
and $q$ is the so-called nonextensivity parameter which measures the temperature 
fluctuations \cite{q_Tsallis} in the system as: $q-1 = Var(T)/<T>^{2}$.
 The values of $q$  lie between $1 < q < 4/3$. Using the relation $ 1/(q-1) = n$, 
Eq.~\ref{Tsallis_eqn} takes the form

\begin{eqnarray}
E\frac{d^3N}{dp^3}  & = &  C_{n}\left(1 + \frac {E}{nT}\right)^{-n}, \nonumber \\
                    & = &  C_{n}\left(1 + \frac {m_T}{nT}\right)^{-n}.
\label{Tsallis}
\end{eqnarray}
 Here $C_{n}$ is the normalization constant and $m_T=\sqrt{p_T^2 + m^2}$. Larger values of $q$ 
correspond to smaller values of $n$ 
describing a system away from thermal equilibrium. In terms of QCD, smaller values of
$n$ imply dominant hard point-like scattering. Phenomenological studies suggest that, 
for quark-quark point scattering, $n\sim$ 4 \cite{BlankenbeclerPRD12, BrodskyPLB637}, 
and when multiple scattering centers are involved $n$ grows larger and can go up to 
20 for proton $p_T$ spectra.

  If there exists a transverse flow of particles in a co-moving frame or system, then 
we can replace the energy by the following four vector form 
\begin{eqnarray}
 E = v^{\mu}p_{\mu} = \gamma(m_{T} - \vec{\beta} . \vec{p_{T}}),
\label{Four_vector}
\end{eqnarray}
 where the factor $\gamma = 1/\sqrt{1-\beta^{2}}$, $v^{\mu} = \gamma(1, \vec{\beta}, 0)$ 
and $p_{\mu} = (m_{T}, -\vec{p_{T}}, 0)$ are four-velocity and four-momentum 
of particles in central rapidity region. 
 Assuming, $\vec{\beta}$ and $\vec{p_T}$ to be collinear and denoting the average transverse velocity of
the system by $\beta$, the Tsallis distribution in Eq.~\ref{Tsallis} takes the form 

\begin{eqnarray}
 E\frac{d^3N}{dp^3}  =  C_{n}\left(1 + \frac {\gamma (m_{T} - \beta p_{T})}{nT}\right)^{-n}.
\label{Tsallis_flow}
\end{eqnarray}
 
 There have been many attempts to use Tsallis distributions in hydrodynamical models  
[e.g. \cite{arXiv0912.0993v3,zebo}]. These formalisms are used to explain the hadronic spectra 
produced in heavy ion collisions in the $p_T$ range of 0 to 3 GeV/$c$.
  Our goal is to find an analytical fit function which works in a wider $p_T$ range. 
We propose a modified Tsallis formula given by   
\begin{eqnarray}
E\frac{d^3N}{dp^3}  =  C_{n}\left({\rm exp}\left(\frac {-\gamma \beta p_{T}} {nT}\right) + \frac {\gamma m_{T}}{nT}\right)^{-n}.
\label{Tsallis_mod}
\end{eqnarray}
  The low and high $p_T$ limits of this formula are given by  
\begin{eqnarray}
\label{boltz}
 E\frac{d^3N}{dp^3}
  &\simeq \exp\left(\frac {-\gamma(m_{T} - \beta p_T)} {T}\right) \, \, \,  {\rm for}\,\,\, p_{T} \rightarrow 0 \\
  &\simeq \left(\frac  {\gamma m_{T}} {nT} \right)^{-n}  \,\,\,\,\,\,{\rm for}\,\,\, p_{T} \rightarrow \infty. 
\label{power}
\end{eqnarray}  
  Thus at low $p_T$, it represents a thermalized system with collective flow and at high $p_T$
it becomes a power law which mimics "QCD inspired" a quark interchange model \cite {HAGEFACT}.

\begin{figure*}
\begin{center}
\includegraphics[width=0.62\textwidth]{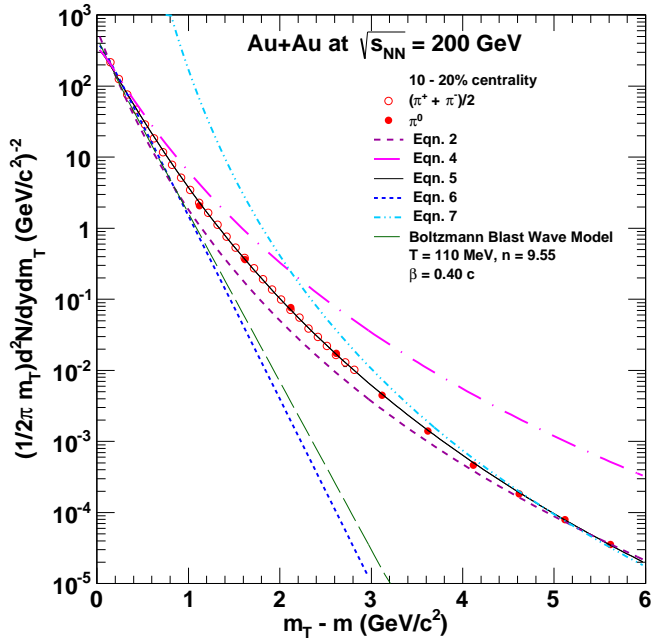}
\caption{(Color online) Comparison of different fit functions. 
  The modified Tsallis function fitted on charged and neutral pions measured in 10-20 \% central 
Au+Au collisions at $\sqrt{s_{NN}}$ = 200 GeV. The same set of parameters is used to obtain other functions 
described in the text.}
\label{Compare_TandBoltzman} 
\end{center} 
\end{figure*}

 Figure~\ref{Compare_TandBoltzman} gives a comparison of different fit functions namely the Tsallis  
by Eq.~\ref{Tsallis}, Tsallis including radial flow by Eq.~\ref{Tsallis_flow}, modified Tsallis by 
Eq.~\ref{Tsallis_mod}, Boltzmann distribution including flow by Eq.~\ref{boltz} and power law by Eq.~\ref{power}
along with the blast wave model from Ref~\cite{Schnedermann} using Boltzmann distribution.
  The modified Tsallis function is fitted on charged and neutral pions measured in 
10-20 \% central Au+Au collisions at $\sqrt{s_{NN}}$ = 200 GeV to get the fit parameters which
are then used to calculate all other functions given in the figure. 
 The normalizations for all functions are arbitrary as our aim is to compare the shapes of different 
functions. 
 On comparing the modified Tsallis with the original Tsallis we can study the effect of parameter 
$\beta$ which enhances the contribution in intermediate $p_T$ range. 
 In the low $p_T$ range the modified Tsallis tends to an exponential distribution given by Eq.~\ref{boltz}
and to a power law in high $p_T$ range. Thus it preserves the well known property of the Tsallis distribution 
which at low $p_T$ tends to an exponential of form Eq.~\ref{boltz} (with $\beta=0$) and at high 
$p_T$ is consistent with a power law.
 Another interesting thing we note is, both the blast wave model and the distribution given by Eq.~\ref{boltz} 
including average radial flow velocity are similar describing the data in low $p_T$ range.
 
 The Tsallis including radial flow by Eq.~\ref{Tsallis_flow} is also shown in the figure. This does not give
the both low and high $p_T$ limits for the same set of parameters. It is possible to fit the data with this
function using a very large value of $n$. It is expected that the blast wave model 
using Tsallis \cite{arXiv0912.0993v3,zebo} has similar behavior which is indicated by large values of 
$n$ (smaller values of $q-1$) obtained in their work.

\begin{table*}
\caption{Particles used in this analysis, $p_{T}$ and rapidity
 ranges in $p+p$ and Au+Au collisions at $\sqrt{s}$ = 200 GeV.} 
\begin{center}
\label{refer1}
\begin{tabular}{cccc}
\hline
Particle      & $p_T$ range (GeV/$c$) &  Rapidity range  &  Reference              \\    
\hline 
 $\pi^{0}$    & 0.6 - 6.0        & $|y|<0.35$     & PHENIX  \cite{PPG063, PPG080}    \\
     
 $\pi^{\pm}$  & 0.3 - 2.6        & $|y|<0.35$     & PHENIX  \cite{PPG030, PPG026}    \\

 $K^{\pm}$    & 0.4 - 1.8        & $|y|<0.35$     & PHENIX  \cite{PPG030, PPG026}    \\

 $p$          & 0.47 - 6.0       & $|y|<0.55$     & STAR  \cite{ppproton, auauproton} \\

 $\Lambda$    & 0.35 - 4.75      & $|y|<1.0$     & STAR \cite{ppbaryon, auaubaryon}  \\

 $\Xi^{-}$    & 0.67 - 3.37      & $|y|<0.75$     & STAR \cite{ppbaryon}  \\
             & 0.85 - 5.0        & $|y|<0.75$    & STAR \cite{auaubaryon}  \\
           
\hline
\end{tabular}
\end{center}
\end{table*}

  We test the formula (Eq.~\ref{Tsallis_mod}) using a large amount of data on transverse momentum 
spectra measured in $p+p$ and Au+Au collisions at  $\sqrt{s_{NN}}$ = 200 GeV, which are 
listed in Table~\ref{refer1} along with their references.
 Here we use all the data from PHENIX ($|y| < 0.35 $) \cite{PPG030, PPG063, PPG026, PPG080} and 
STAR ($|y| < 0.5 $, $|y| < 0.75 $) \cite{ppproton, ppbaryon, auauproton, auaubaryon} experiments.
The errors on the data are quadratic sums of statistical and uncorrelated systematic errors wherever available. 
  The particle spectra are studied as a function of system size using
p+p collisions and Au+Au collisions of four centralities; 10-20 \%, 20-40 \%, 40-60 \% and 60-80 \%.

 We employ the $\chi^2$ method to fit the data and Table~\ref{refer2} gives the $\chi^2$ and number of 
degrees of freedom.
  During our fits we consider a large $p_T$ range and do not assume that the parameters 
for all particles are the same in a particular collision system. 
 The values of $n$ are decided by the initial hard scatterings which involves point like qq  
or multiple scattering centers and is different for different particles and they depend 
on energy of the colliding system \cite{IJMPA}. 

 The three parameters $T$, $n$ and $\beta$ are not completely independent of each other and
effect of one parameter can be absorbed in the other to certain extent. 
  Thus in our fit procedure first we fit the measured spectra of pions, kaons, protons, $\Lambda$ and $\Xi$ 
in $p+p$ collisions using the modified Tsallis distribution (Eq.~\ref{Tsallis_mod}) with the same temperature
to obtain different values of $n$ (Table~\ref{Table_n}) for different particles. The values of $\beta$ are small for 
$p+p$ collisions as is expected.
  The spectra of different particles in Au+Au collisions for all centralities are fitted to obtain
$\beta$ and $T$ as a function of system size. 
  When going from p+p collisions to Au+Au collisions, the initial hard scattering is assumed 
to be the same and hence the value of $n$ for each particle is kept fixed to values obtained from 
$p+p$ collisions. At high $p_T$ the shape of particle spectra in Au+Au collisions remains a power 
law without a noticeable increase in value of $n$ obtained from $p+p$ spectra so this assumption
is good and helps us studying the behavior of the other two parameters as we increase the 
system size.

\begin{table}
\caption{Values for $n$, obtained from the fits in $p+p$ system at $\sqrt{s}$ = 200 GeV.}
\label{Table_n}
\begin{center}
\begin{tabular}{cc}
\hline
Particle     &   Values of $n$ \\
\hline
\hline
$\pi$        &  9.55 $\pm$ 0.13 \\
$k$          &  8.61 $\pm$ 0.24 \\
$p$          &  11.95 $\pm$ 0.10\\
$\Lambda$    &  13.24 $\pm$ 5.36\\
$\Xi$        &  13.24 $\pm$ 5.18 \\

\hline
\end{tabular}
\end{center}
\end{table}

\begin{table}
\caption{$\chi^{2}$/ndf values for the fits in $p+p$ and Au+Au collisions at $\sqrt{s_{NN}}$ = 200 GeV.}
\label{refer2}
\begin{center}
\begin{tabular}{cccccc}
\hline
Particle     &   \multicolumn{4}{c} {$\chi^{2}/NDF$  values for fits} \\
\hline
             & $p+p$ system &        &    Au+Au system & & \\
             &          & 10 - 20\%  & 20 - 40\% & 40 - 60\% & 60 - 80\%\\
\hline
$\pi$        &  28.2/27 &  11.7/35   &  34.2/35 &  31.4/35 &  24.1/35 \\
$k$          &  7.4/10  &  20.2/13   &  30.0/13 &  20.6/13 &  7.0/13  \\
$p$          &  21.8/16 &  11.3/21   &  12.7/21 &  10.5/21 &  8.1/21  \\
$\Lambda$    &  14.2/18 &  14.3/14   &  11.5/14 &  3.6/14  &  17.0/14 \\
$\Xi$        &  4.5/ 8  &  10.7/10   &  3.9/10 &   7.2/10  &  0.6/ 3  \\

\hline
\end{tabular}
\end{center}
\end{table}

\section{Results and Discussions}


\begin{figure*}
\begin{center}
\includegraphics[width=0.98\textwidth]{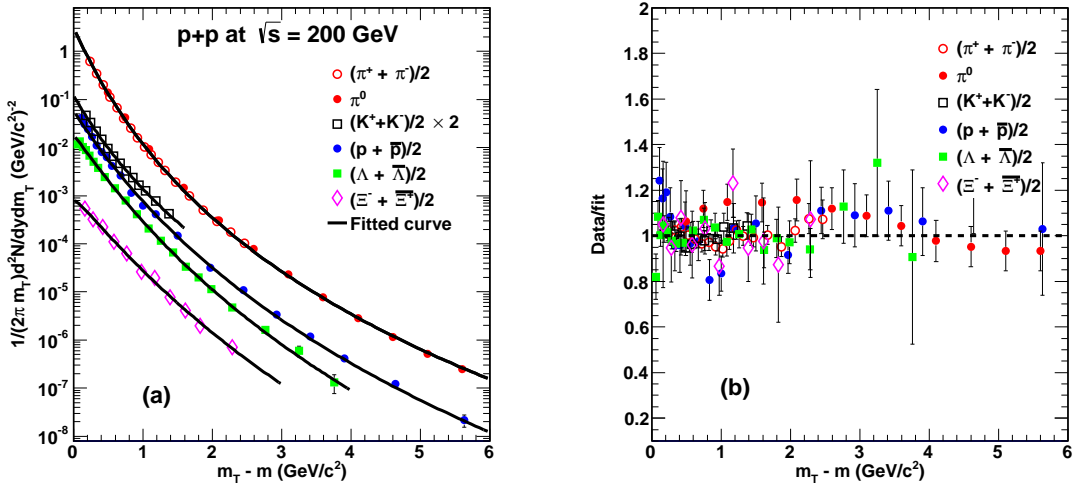}
\caption{(Color online) (a) The invariant yields of pions ($(\pi^{+}+\pi^{-})/2$, $\pi^{0}$) \cite{PPG030, PPG063}, 
kaons ($(K^{+}+K^{-})/2$)\cite{PPG030},  protons ($(p+\bar{p})/2$) \cite{ppproton}, 
$\Lambda$ ($(\Lambda+\bar{\Lambda})/2$) \cite{ppbaryon} and 
$\Xi$ ($(\Xi^{-}+\bar{\Xi^{+}})/2$) \cite{ppbaryon}
as a function of $m_{T}$ for $p+p$ collision at $\sqrt{s}$ = 200 GeV. 
The solid lines are the modified Tsallis distribution (Eq.~\ref{Tsallis_mod}).
(b) Shows the Data/fit.}
\label{pphadron} 
\end{center} 
\end{figure*}

\begin{figure*}
\begin{center}
\includegraphics[width=0.98\textwidth]{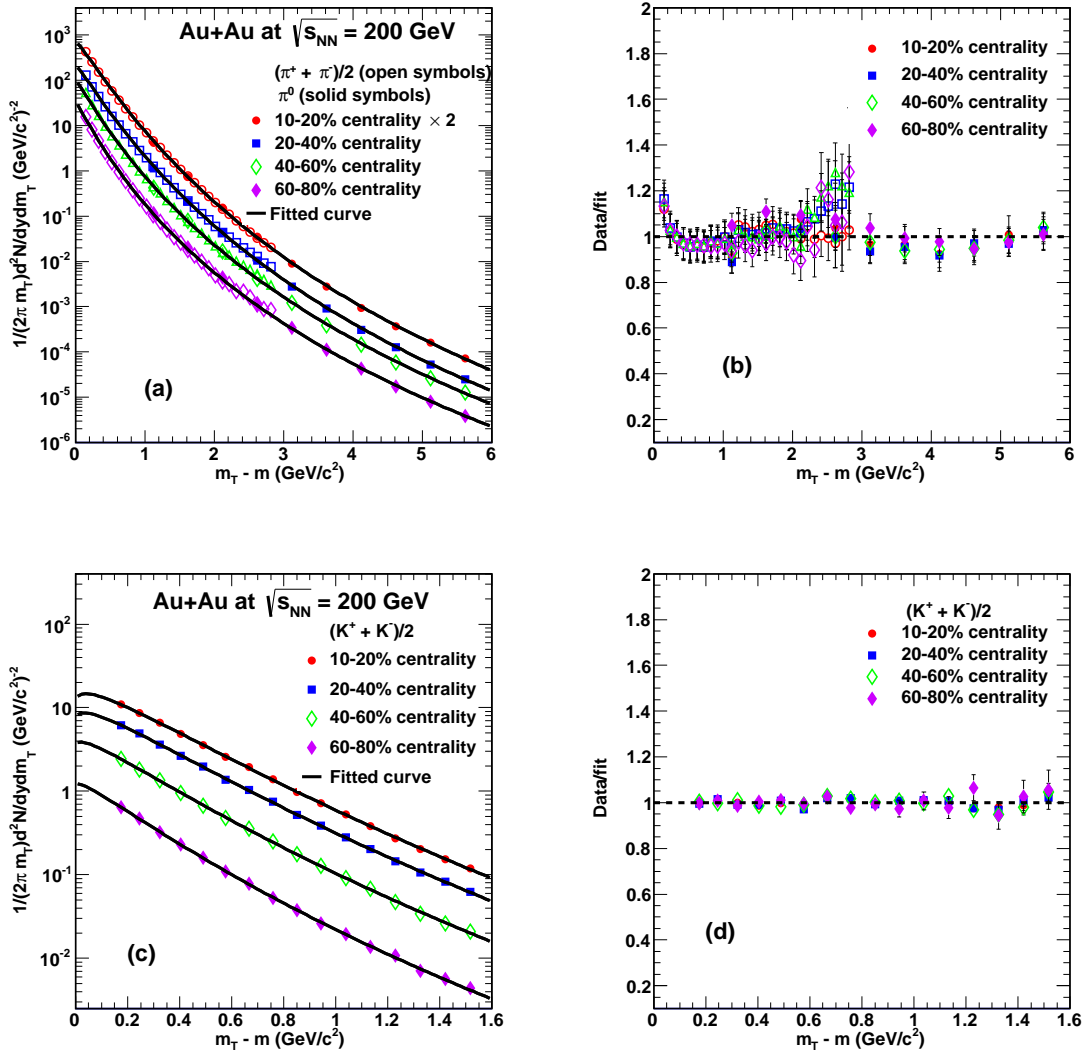}
\caption{(Color online) The invariant yields of (a) pions ($(\pi^{+}+\pi^{-})/2$ \cite{PPG026}, $\pi^{0}$ \cite{PPG080}) and
(c) kaons ($(K^{+}+K^{-})/2$)\cite{PPG026}
as a function of $m_{T}$ for Au+Au collision at $\sqrt{s_{NN}}$ = 200 GeV. 
The solid lines are the modified Tsallis distribution (Eq.~\ref{Tsallis_mod}).
 (b) and (d) show the Data/fit.}
\label{auau_pion_kaon} 
\end{center} 
\end{figure*}

\begin{figure*}
  \begin{center}
\includegraphics[width=0.98\textwidth]{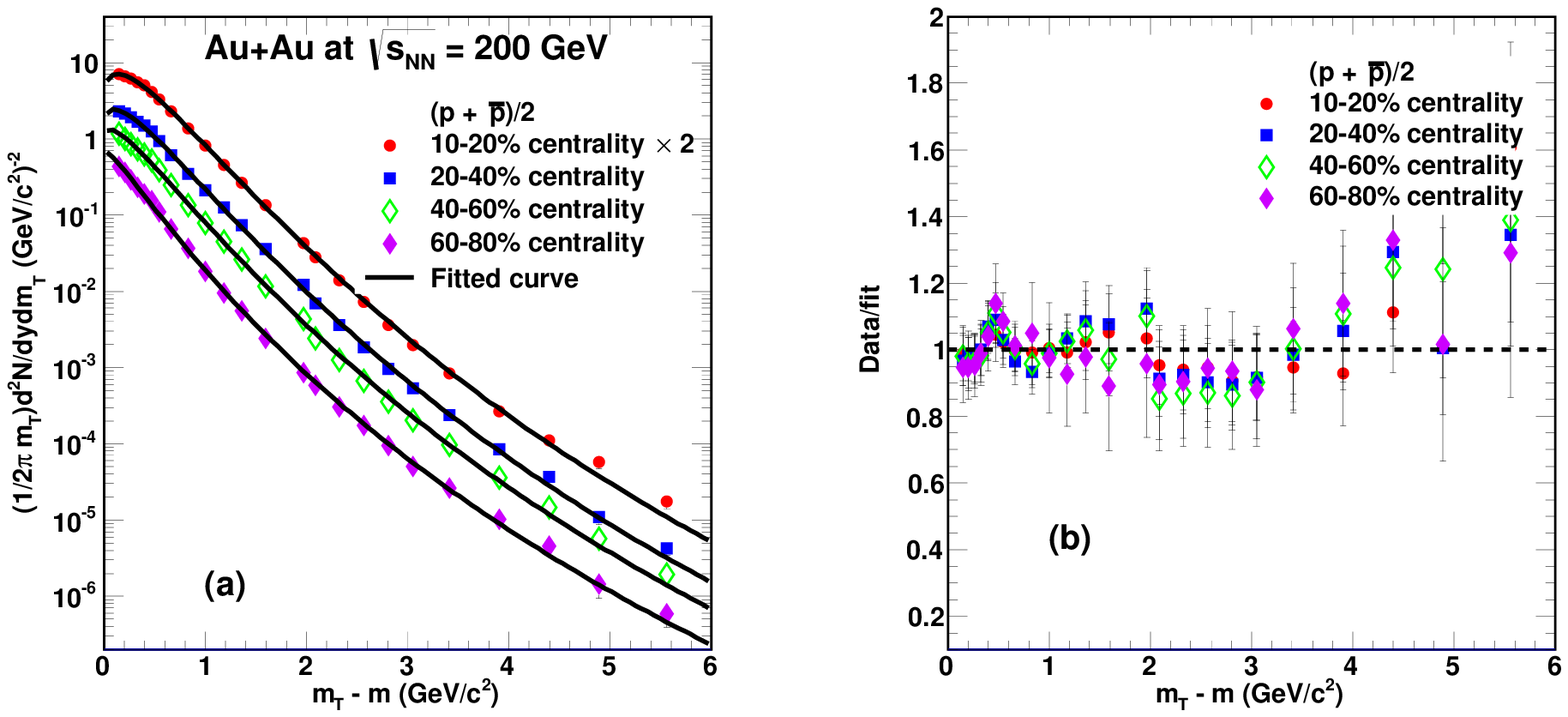}
\caption{(Color online)(a) The invariant yield of proton ($(p+\bar{p})/2$) \cite{auauproton}
as a function of $m_{T}$ 
in Au+Au collisions at $\sqrt{s_{NN}}$ = 200 GeV for different centralities. 
The solid lines are the modified Tsallis distribution (Eq.~\ref{Tsallis_mod}).
(b) Shows the Data/fit.} 
\label{auauproton} 
\end{center} 
\end{figure*}

\begin{figure*}
\begin{center}
\includegraphics[width=0.98\textwidth]{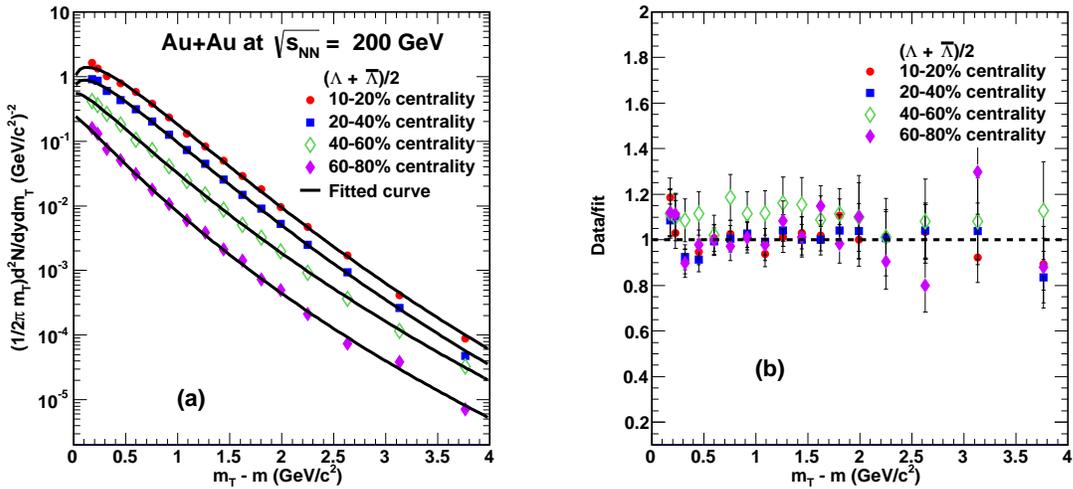}
\caption{(Color online) (a) The invariant yield of $\Lambda$ ($(\Lambda+\bar{\Lambda})/2$) \cite{auaubaryon} 
as a function of $m_{T}$ in Au+Au collisions at $\sqrt{s_{NN}}$ = 200 GeV for different centralities.
The solid lines are the modified Tsallis distribution (Eq.~\ref{Tsallis_mod}).
(b) Shows the Data/fit.}  
\label{auauLambda}
\end{center}
\end{figure*}

\begin{figure*}
\begin{center}
\includegraphics[width=0.98\textwidth]{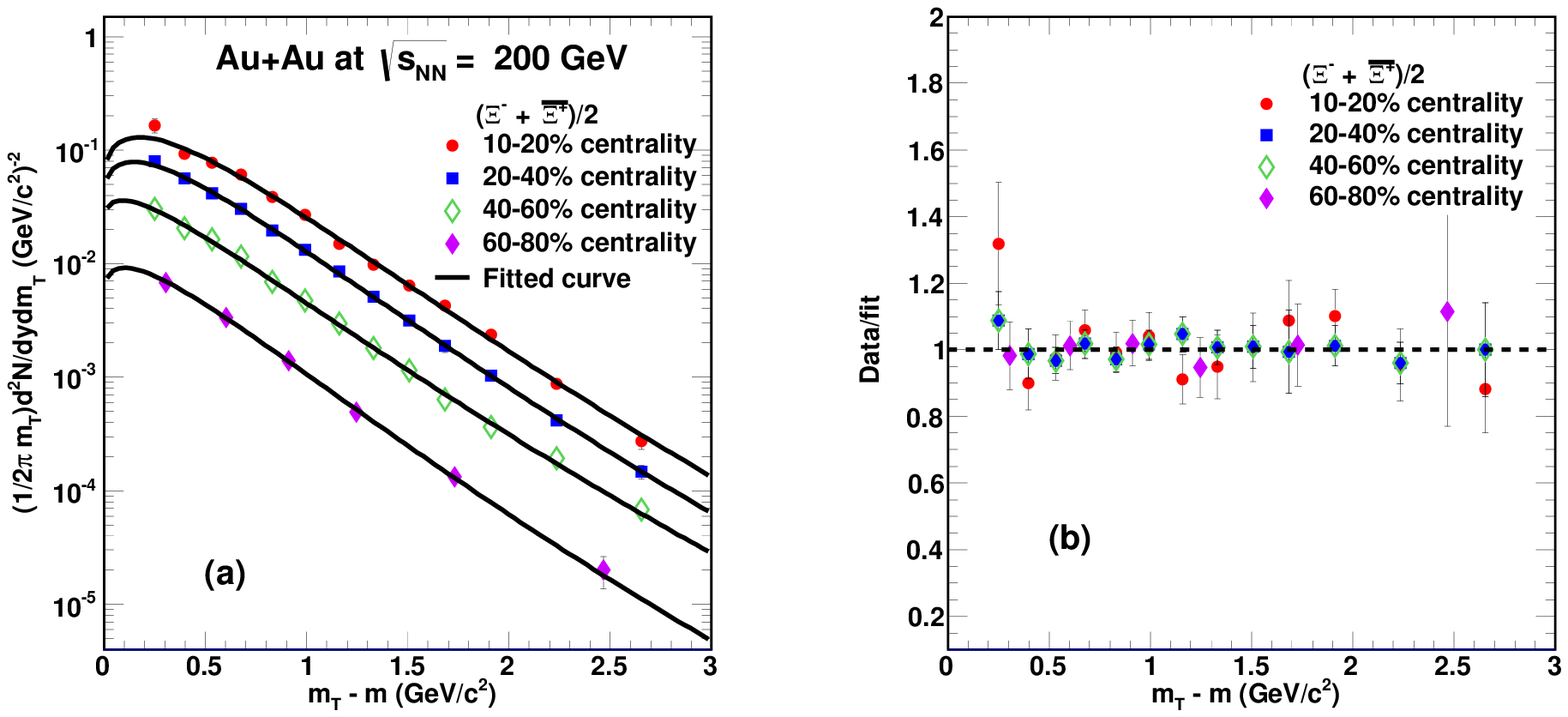}
\caption{(Color online)(a) The invariant yield of $\Xi$ ($(\Xi^{-}+\bar{\Xi^{+}})/2$) \cite{auaubaryon}
as a function of $m_{T}$ in Au+Au collisions at 
$\sqrt{s_{NN}}$ = 200 GeV for different centralities. 
The solid lines are the modified Tsallis distribution (Eq.~\ref{Tsallis_mod}).
(b) Shows the Data/fit.} 
\label{auauXi}
\end{center}  
\end{figure*}

\begin{figure}
\begin{center}
\includegraphics[width=1.0\textwidth]{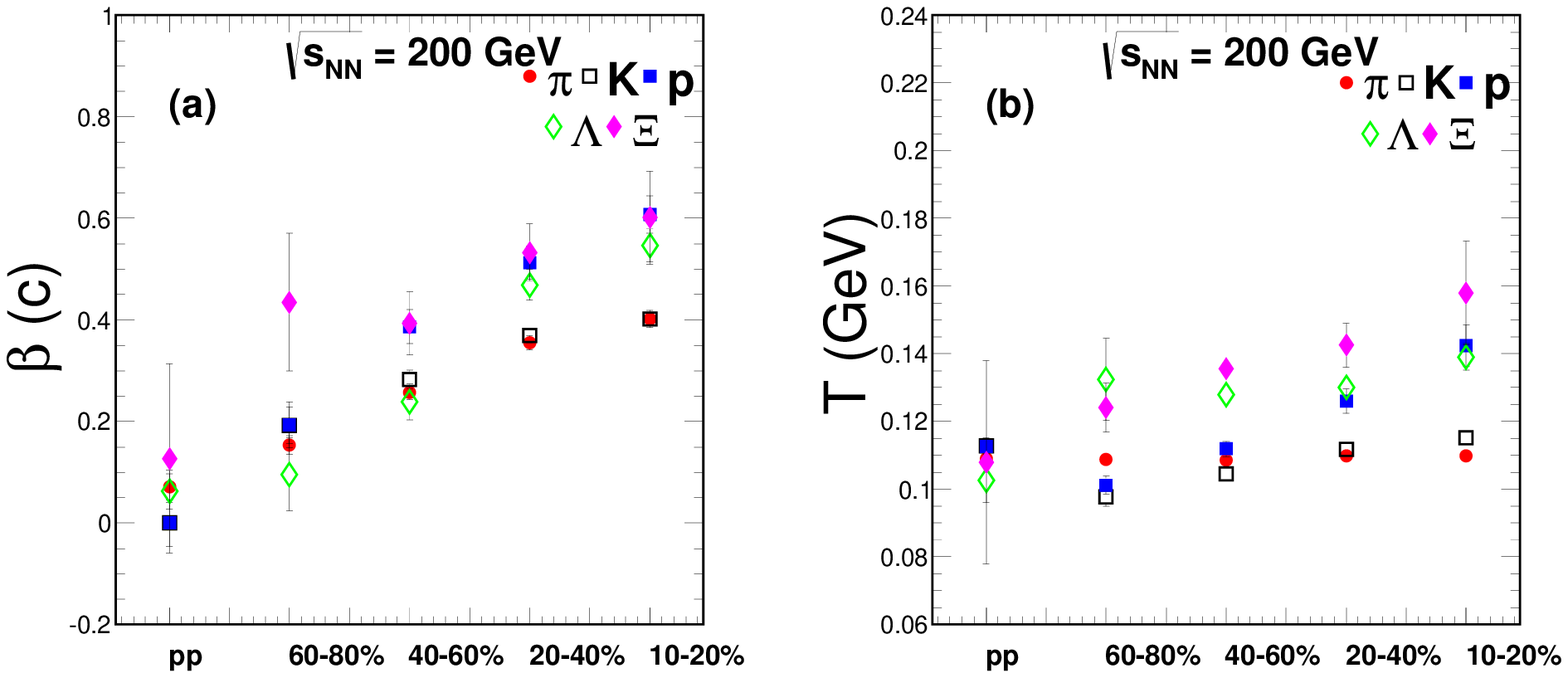}
\caption{(Color online) The variation of modified Tsallis parameters
(a) the transverse flow velocity $\beta$, 
(b) the freeze-out temperature $T$ as a function of collision system size 
at $\sqrt{s_{NN}}$ = 200 GeV for pions, kaons, protons,
$\Lambda$ and $\Xi$.}

\label{auauparameter} 
\end{center} 
\end{figure}


  Figure~\ref{pphadron}(a) shows the transverse mass spectra of pions \cite{PPG030, PPG063}, 
kaons \cite{PPG030}, protons, $\Lambda$ and $\Xi$ \cite{ppproton} in $p+p$ collisions at 
$\sqrt{s}$ = 200 GeV fitted with the modified Tsallis function [Eq.~\ref{Tsallis_mod}]
with the same temperature to obtain different values of $n$. 
The fit quality is good for all particles in all $p_T$ range (0.3 to 6 GeV/$c$) shown by 
ratio of data points to the fit function plotted in Fig.~\ref{pphadron}(b).

  Figures~\ref{auau_pion_kaon} (a) and (c) show transverse mass spectra of pions
\cite{PPG026,PPG080} and kaons \cite{PPG026}, respectively in Au+Au collisions 
at $\sqrt{s_{NN}}$ = 200 GeV for all four centralities fitted with the modified Tsallis function. 
Here the values of $n$ for pions and kaons are fixed to the values obtained from $p+p$ collisions.  
 Figures~\ref{auau_pion_kaon}(b) and (d) give the ratio of data points to the fit function for pions and kaons respectively
which show the quality of fit. One can notice that the modified Tsallis gives a very good 
fit for pions in all centralities of Au+Au collisions from very low $p_T$ to high $p_T$. 
 The kaon spectrum,  available only at low $p_T$, is also well described.

  Figure~\ref{auauproton}(a) shows transverse mass spectra of protons \cite{auauproton} 
in Au+Au collisions at $\sqrt{s_{NN}}$ = 200 GeV for different centralities.
Figure~\ref{auauproton}(b) shows the ratio of data points to the fit function. Except for the last two points for most central collisions, 
the data of all centralities are well reproduced. 
 Figure~\ref{auauLambda} is same as Fig.~\ref{auauproton} 
but for $\Lambda$ \cite{auaubaryon}. Figure~\ref{auauXi} is as Fig.~\ref{auauproton} 
but for $\Xi$ \cite{auaubaryon}. The modified Tsallis function gives very good quality of 
fit for all hadrons measured in Au+Au collisions at all centralities. 
The $\chi^{2}/NDF$ values for fits in $p+p$ and Au+Au system are listed in Table~\ref{refer2}.

  Figure~\ref{auauparameter} shows the variation of Tsallis parameters:
(a) the transverse flow velocity $\beta$ and 
(b) the freeze-out temperature $T$ 
for pions, kaons, protons,
$\Lambda$ and $\Xi$ for $p+p$ and different centralities of Au+Au collisions 
at $\sqrt{s_{NN}}$ = 200 GeV. The errors on the parameters include the error on data
and possible correlations among the parameters.
  The power $n$ for each particle are obtained by its $p_T$ spectrum in $p+p$ collisions.
    The flow velocities ($\beta$) for all particles are very small in $p+p$ collisions 
and increases with centrality in Au+Au collisions which is expected.
 For central collisions, there is clear separation of flow velocity between mesons and 
baryons showing a dependence on number of constituent quarks. 
For the most central Au+Au collisions, the values of $\beta$ for protons is 
around 0.6 and for pions it is 0.4. Thus,  when analyzing the spectra it is more important to 
group the particles as baryons and mesons than to apply any other criteria e.g. one based on 
strangeness.
  The behavior of freeze-out temperatures $T$ can also be grouped into mesons and baryons. 
With increasing system size, $T$ is almost constant for pions and increases weakly for kaons.
 For baryons the increase of temperature is more rapid as compared to mesons. 
  There is also a clear separation of temperature between baryons and mesons towards most 
central Au+Au collisions. It means that baryons freezeout earlier than mesons. 


\section{Conclusion}
  In this work we present a modified Tsallis function to describe the hadron spectra
in heavy ion collisions. 
 The function is analytic and which makes it convenient to use for fitting experimental 
hadronic spectra. 
 The additional parameter we introduced accounts for transverse flow at low $p_T$.
 In low $p_T$ range the modified Tsallis tends to an exponential distribution with radial flow 
and to a power law in high $p_T$ range. Thus it preserves the well known property of 
the Tsallis distribution which at low $p_T$ tends to an exponential form with zero flow velocity and
a power law at high velocities.
  With this modified Tsallis function we study the spectra of pions, kaons, protons, 
$\Lambda(1115)$, and $\Xi^{-}$ as a function of system size at $\sqrt{s_{NN}}$ = 200 GeV. 
 This new analytical function gives a very good description of both meson and baryon 
spectra in all centralities of Au+Au collisions in terms of parameters namely, temperature,
power and transverse flow. 
  We observe that the flow velocity extracted for all particles increases with centrality 
in Au+Au collisions. There is a clear separation of flow  between mesons and baryons in 
the most central Au+Au collisions showing a dependence on number of constituent quarks.
  The behavior of freeze-out temperatures $T$ can also be grouped into mesons and baryons. 
 For baryons the increase of temperature is more rapid than for mesons. The baryons
in general freeze out earlier than the mesons.

\ack
We acknowledge the financial support from Board of Research in Nuclear 
Sciences (BRNS) for this project.

\end{document}